# Network Slicing to Enable Scalability and Flexibility in 5G Mobile Networks

P. Rost[1], C. Mannweiler[1], D. S. Michalopoulos[1], C. Sartori[1], V. Sciancalepore[2], N. Sastry[3], O. Holland[3], S. Tayade[4], B. Han[4], D. Bega[5,6], D. Aziz[1], and H. Bakker[1]

[1]Nokia Bell Labs Germany, [2]NEC Laboratories Europe, [3]King's College London, [4]University of Kaiserslautern, [5]Univeristy Carlos III of Madrid, [6] IMDEA Networks Institute

**Abstract:** We argue for network slicing as an efficient solution that addresses the diverse requirements of 5G mobile networks, thus providing the necessary flexibility and scalability associated with future network implementations. We elaborate on the challenges that emerge when we design 5G networks based on network slicing. We focus on the architectural aspects associated with the coexistence of dedicated as well as shared slices in the network. In particular, we analyze the realization options of a flexible radio access network with focus on network slicing and their impact on the design of 5G mobile networks. In addition to the technical study, this paper provides an investigation of the revenue potential of network slicing, where the applications that originate from such concept and the profit capabilities from the network operator's perspective are put forward.

## I. Introduction

Future mobile networks will be subject to a manifold of technical and service requirements with respect to throughput, latency, reliability, availability, as well as operational requirements such as energy-efficiency and cost-efficiency. These requirements stem from an increasing diversity of services carried by the mobile network as well as novel application areas such as Industry 4.0, vehicular communication, or smart grid. In order to provide cost- and energy-efficient solutions, it is necessary to avoid a largely segmented solution space with deployments of individual mobile network solutions for each use case. Hence, there is the need for a flexible and scalable mobile network. Thereby, *flexibility and scalability* go hand in hand and make sure that the mobile network can be appropriately adopted to the network environment of a particular use case, e.g., available bandwidth, transport network, or access point density. Furthermore, the actual quantitative technical requirements may differ significantly, e.g., while packet error rates of $10^{-4}$ are acceptable in a mobile broadband system, industrial use cases require significantly lower packet *and* frame error rates, in particular if latency constraints must be met [1].

### A. *Definition of Network Slices*

In order to cope with the above requirements, the concept of network slicing has been proposed as a means for providing better resource isolation and increased statistical multiplexing [2]. The Next Generation Mobile Network Alliance (NGMN) defines network slicing as a concept for running multiple logical networks as independent business operations on a common physical infrastructure [2]. Each network slice represents an independent virtualized end-to-end network and allows operators to run different deployments based on different architectures in parallel. In the following, the term *network slice* refers to a specific instance of such a logical network (instantiated according to a pre-defined *network slice blueprint*).

A network slice as logical end-to-end construct is self-contained, having customized functions including also those in the user equipment (UE), and using network function chains for delivering services to a given group of devices. Employing network slicing in 5G networks engenders a number of challenges, in part due to difficulties in virtualizing and apportioning the Radio Access Network (RAN) into different slices, as discussed in the ensuing sub-section.

### B. Design Challenges

In the following, we provide a detailed explanation of the potential challenges associated with the implementation of network slicing in future networks.

**Granularity constraints in spectrum and radio-level resource sharing**: Unlike fixed network slices which can be scaled up by adding more hardware resources, RAN slicing quickly runs into a *physical* constraint: The limited availability of spectrum. This limitation is deteriorated if dedicated carriers are assigned to individual slices, since such approach does not leverage the network's potential for multiplexing gains.

**Radio Access Technology (RAT) heterogeneity and spatial diversity**: It is expected that 5G will incorporate several kinds of RATs and air interfaces, each with different capabilities and needs. General-purpose infrastructure providers will need to carefully plan and apply different technologies to serve diverse tenant needs. Yet, it may be infeasible to satisfy the needs of each application at any location. For instance, Tactile Internet may require careful positioning of resources to minimize latency. In another example, an Industrial control network might *have* to use a certain computational resource in a given location for security reasons.

**Managing information exposure and sharing constraints:** Different flavors of network slices can be defined based on the extent of network elements that are shared, e.g., whether only the PHY is shared, whether the MAC layer is shared, or even whether the complete RAN is shared. The more the information that can be provided by the infrastructure about the shared parts to the network slice, the more efficient the slice can be operated. However, exposing information also creates new potential security vulnerabilities between infrastructure-providers and their clients (also known as "tenants" [2]), as well as between tenants themselves. Security requirements of specific tenant applications, such as traffic associated with emergency services, or machine control (e.g., remote surgery or vehicular control), could put constraints on how the slices are partitioned, or even prevent network slices to co-exist and thus share the same hardware at all.

**Transparency of network slicing**: A major question is whether a slice can be extended all the way to the UE. That is, whether the definition of the slice will be transparent to the UE, or whether the UE will be aware of the network slice. A slicing-aware UE may open up new possibilities, e.g., a simplification of multi-slice connectivity. However, it also creates new challenges for network slices, e.g., UE mobility may need to be handled by the slice provider as part of the slice setup and maintenance.

**Network slice requests brokerage:** Network slicing in 5G networks enables a new ecosystem in which different tenants issue requests to an infrastructure provider for acquiring network slices. Since spectrum is a scarce resource for which overprovisioning is not possible, applying an "always accept" strategy for all incoming requests is not feasible. This calls for novel algorithms and solutions to allocate network resources among different tenants, allowing an infrastructure provider to accept or reject network slice requests with the objective of maximizing the overall utility.

### C. Network Slicing Applications and Profitability

This section highlights the major applications where the slicing concept is expected to play a key role in future networks, along with a profitability assessment as seen through the lens of the operator.

#### 1) Slicing Applications: Smart Factory and the Tactile Internet

Two exemplary applications for network slicing are "smart factory" industrial communications and the "Tactile Internet". In both cases, wireless communication conveys force (or "kinaesthetic") information to a client, and in the Tactile Internet case especially touch sensations such as texture might be conveyed. The purpose of these applications is to achieve the touching or manipulation of remote real or virtual objects by a human or machine. If kinaesthetic information is conveyed to a machine client, the latency requirement might correspond to the challenging 1ms in 5G. For human clients, this is relaxed to around 5ms, or more than 100ms for tactile information alone conveyed to humans. Both applications also require extremely high reliability and security requirements, noting the mission-critical characteristics associated with them.

Network slicing can address the latency, reliability and security requirements of these applications. Referring to the remote surgery example shown in Figure 1, virtualization allows the instantiation of network elements at appropriate locations for the communication to proceed as close to a direct path as possible, reducing propagation delay hence latency. The instantiation of virtualized elements collectively forming network slices allows multiple instances of such applications to viably share available computational and other resources end-to-end, making virtualization viable from a management point of view. Slicing also assists reliability through the reservation of hardware and other resources as distinct slices, even in some cases potentially down to spectrum resources. Security can benefit through slicing, e.g., tenant isolation and "sandboxing" capabilities. Furthermore, slices may only be operated locally within a factory in order to ensure data privacy while its operation is coordinated with slices operated by public MNOs offering Internet services or specific network functionality such as mobility management.

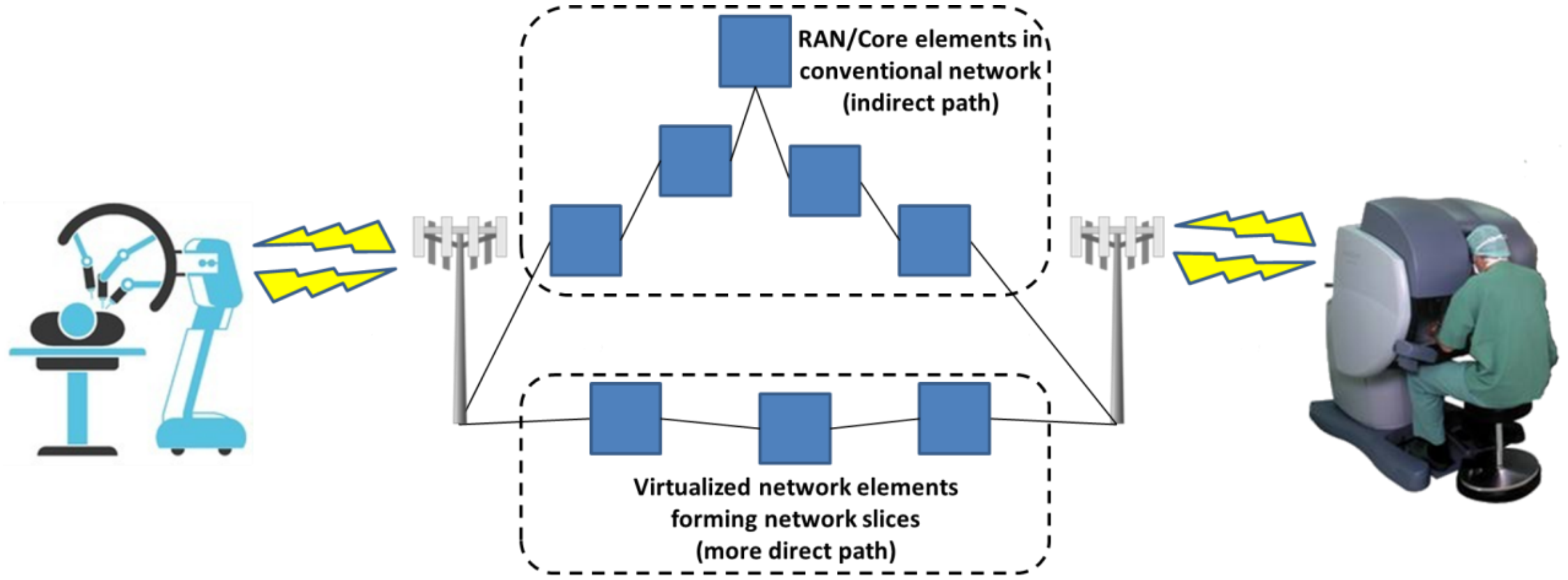

**Figure 1: Virtualized edge network slices achieving a more direct path compared with (fixed) network elements in a Tactile Internet remote surgical operation example.**

*2) Slicing as a Means to Increase Network Revenue*

Beside the flexibilities provided by network slicing, it is also important to demonstrate the economic profit of applying network slicing from the MNO's perspective. The cost in terms of Capital Expenditures (CAPEX) and operation expenses (OPEX) of a network is often much higher in comparison to the revenue expected by the operators. One reason for low revenue is underutilization of the network. According to the KPI requirements, different use cases may have highly specified resource demands. Nevertheless, in the current framework, the operator can only provide the network with an unspecified resource bundle for general utilization. Hence, most of the resources are often reserved for use cases with only slight demands on them, and are thus wasted. With network slicing, the MNOs are able to efficiently analyze the operational cost and revenue generated from the respective slice. According to the analysis, they can allocate different network resource bundles to different slices, which makes the resource management much more structured, flexible, and efficient. As a result, the very same network can be utilized to seamlessly provide more and better services, i.e., generate more revenue without any increase in CAPEX.

Moreover, concepts such as cooperative slicing and inter-operator network sharing can be efficiently implemented by optimizing the network cost model for increasing the overall revenue, and simultaneously providing network scalability. For example, the sliced network of an operator *A* is serving several services and still has few resources unutilized. Hence, the network can implement another slice that requires less resources but more coverage area, and might belong to another operator *B*. The moderated approach of implementing slicing is beneficial for both the operators for providing more services without increasing CAPEX while simultaneously generating revenue from the unutilized resources. Hence, the network provider needs a new algorithm, e.g., based on a threshold-rule, that allows him to decide whether to accept or reject an incoming network slice request while maximizing its revenue.

### D. Related Work

A simplified network slice concept has been exhaustively studied in the literature, wherein dedicated portion of RAN elements are fully reserved to particular services such as an "isolated slice." However, with the advent of advanced network virtualization techniques, the notion of network slicing in 5G has evolved to more flexible sharing, aiming to attain a significant multiplexing gain while still guaranteeing isolation and separation. The Network Virtualization Substrate (NVS) was introduced in [2], allowing the infrastructure provider to control the resource allocation towards each virtual instance of an eNB before each virtual operator customizes scheduling within the allocated resources. In [5], relevant technologies for network slicing are discussed with particular focus on synchronous functions, e.g., multi-dimensional resource management, dynamic traffic steering, and resource abstraction. A particular architecture for network slicing has been introduced and discussed in the context of the 5G NORMA project [6]. Another network slicing solution considering a gateway-based approach is illustrated in [7], wherein a controller provides application-oriented resource abstraction of the underlying RAN. A capacity broker for slice resources has been introduced firstly by the 3rd Generation Partner Project (3GPP) and extensively evaluated in [8] by enabling the on-demand slice resource allocation. The infrastructure provider instantiates a network slice by allocating specific resources to a mobile virtual network operator (MVNO), service providers, and vertical segments for a specified time duration. A study that explores the different options of network sharing based on a centralized broker is provided in [9] considering mobility means for re-directing users to other networks, spectrum transfer policies, and the application of resource virtualization. Finally, [10] discusses a dynamic slicing scheme that flexibly schedules radio resources based on the requested Service Level Agreement (SLA), while maximizing the user rate and applying fairness criteria.

### E. Our Contribution

This work elaborates on the fundamental pillars for an efficient utilization of the concept of network slicing in mobile networks, based on the mobile network architecture framework investigated in the research project 5G NORMA [5]. Particular focus is put on the basic architectural principles for accommodating network slicing in the 5G ecosystem as well as on RAN and core network (CN) aspects. In this regard, we underline the key elements that enable the coexistence of dedicated and shared slices within a common network architecture, and elaborate on the implementation of the notion of network slicing in the RAN and in CN, putting particular emphasis on the concept of Software Defined Mobile network Control (SDMC).

## II. Mobile Network Slicing Architecture

### A. Dedicated and shared sub-slices

Network slices operate on top of a partially shared infrastructure, which is composed of generic hardware resources such as Network Function Virtualization Infrastructure (NFVI) resources, as well as dedicated hardware such as network elements in the RAN. Network functions running on NFVI resources (referred to as virtual network functions (VNFs)) are typically instantiated in a customized manner for each network slice. However, this approach cannot be applied to network functions (NFs) relying on dedicated hardware. Therefore, a key issue for network slicing is the identification and design of common NFs, which are either physical or virtual and which have to be shared by multiple end-to-end slices.

Examples for common NFs include distributed, monolithic eNBs or the radio scheduler in the RAN domain. In the CN domain, candidates for shared VNF instances include Home Subscribe Server (HSS) or mobility management. Generally, three solution groups are discussed with varying levels of common functionality in 3GPP standards [11]: Group A is characterized by a common RAN and completely dedicated CN slices, i.e., independent subscription, session, and mobility management for each network slice handling the UE. Group B also assumes a common RAN, where identity, subscription, and mobility management are common across all network slices, while other functions such as session management reside in individual network slices. Group C assumes a completely shared RAN and a common CN control plane, while CN user planes belong to dedicated slices.

In line with the above grouping considered by 3GPP [11], the framework of the 5G NORMA project [5] introduces dedicated network functions, which together form a dedicated sub-slice and are controlled by the Software-Defined Mobile network Controller (SDM-C). As illustrated in Figure 2, shared network functions are aggregated in common sub-slices that are controlled by the SDM Coordinator (SDM-X), reflecting the fact that these functions have to coordinate and, if necessary, prioritize the Quality of Service (QoS) requirements of multiple slices.

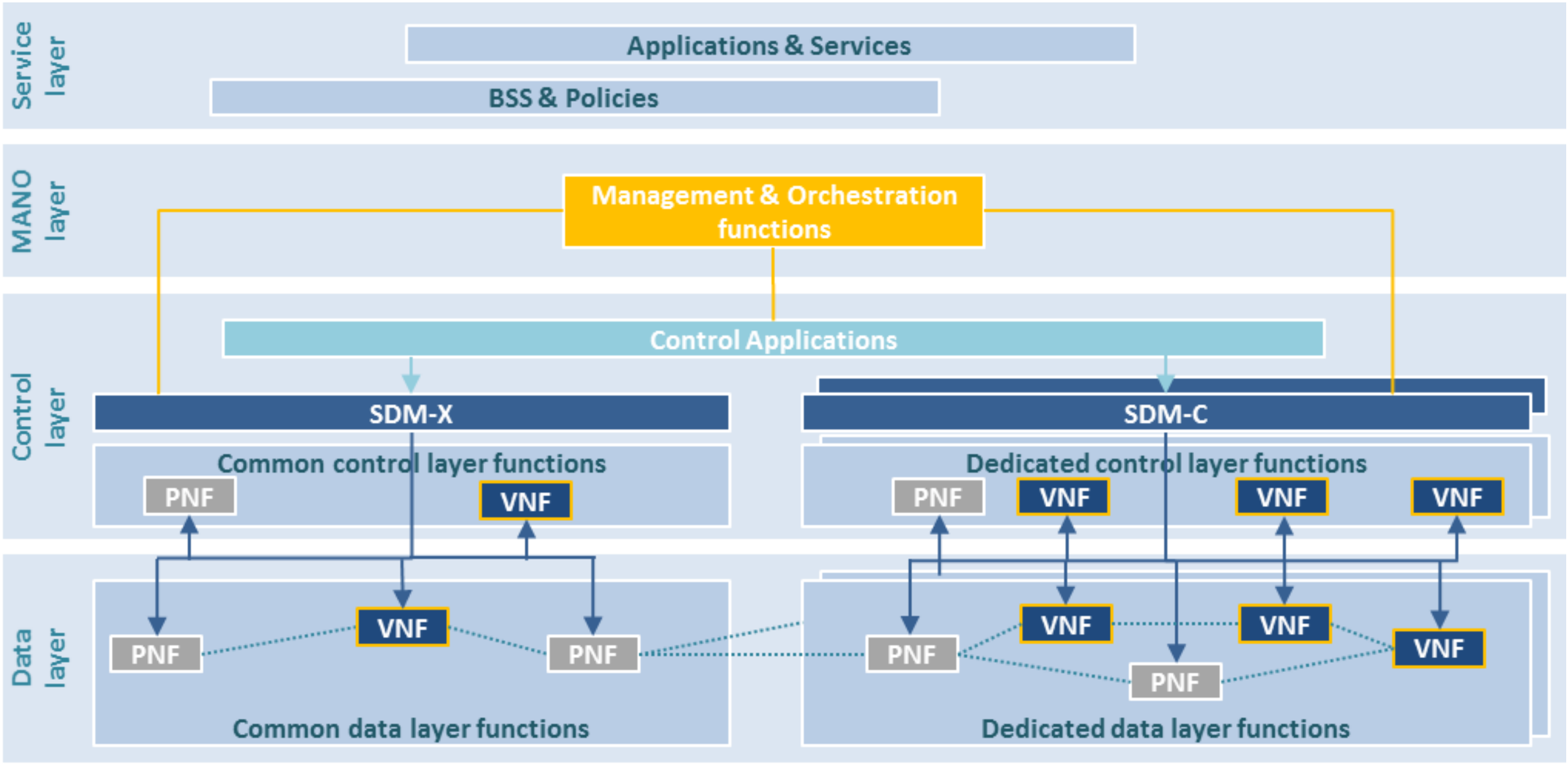

**Figure 2: Combining dedicated and shared sub-slices to form e2e mobile network instances**

## B. *End-to-end network slicing: Common and dedicated network functions*

When sharing NFs and resources between distinct network slices, a central entity in charge of managing and controlling the process is needed, i.e., the SDM-X. This entity ensures to attain high resource efficiency while guaranteeing individual SLAs. Based on the SDMC paradigm, this entity resides on the common control layer; it also includes NFs, either virtual or physical, that the network slices rely on. While a fixed splitting of common NFs (and resources) simplifies the network management and operation, it may lead to an inefficient network utilization. Conversely, dynamic adjustments of common resources might bring multiplexing gains at the expense of less determinism. Hence, the main objective of the SMD-X is to properly administer the trade-off between flexible and static resource assignments, by taking into account sharing policies set by the service provider.

Let us consider the system spectrum as a shared resource pool (divided into several resource blocks (RBs)) fully managed by the SDM-X. The flexibility introduced by the SDM-X enables dynamic and short-term scheduling decisions based on slice requirements. Specifically, the SDM-X facilitates a “masked” view of the shared resource pool towards the network slices. The resource mask is defined as a group of physical RBs dynamically assigned to each network slice. The advantage of such solution relies on the SDM-X channel monitoring phase and on the subsequent dynamic adjustment of slice resource masks, needed to cope with the fast channel dynamics. In a multi-tenancy context [5], a dedicated resource scheduler per tenant may be directly connected to the SDM-X interface, acting as an SDN application. The scheduler uses the slice resource mask and applies its own scheduling policies, while preserving slice isolation constraints. The SDM-X plays a key role in assigning priority to network slices: Different objective functions can be dynamically implemented in order to achieve fairness, maximize spectral efficiency, and mitigate interference.

## III. Implementation of RAN and CN Slicing

### *A. Realization of Network Slicing in CN and RAN*

Figure 3 illustrates how RAN slicing can be realized such that existing and well-proven principles of radio access are utilized. In this regard, the network slice selection function (NS-SF), which is part of the SDM-X concept (c.f. Figure 2), is responsible for selecting the appropriate slice per user. In addition, it configures the RAN-CN interface such that the control and user plane traffic is routed to the accordingly configured functional elements in the CN slice. The user plane anchor (UP-Anchor) is responsible for distributing the traffic according to the configured slice policy, and for encryption with slice-specific security keys.

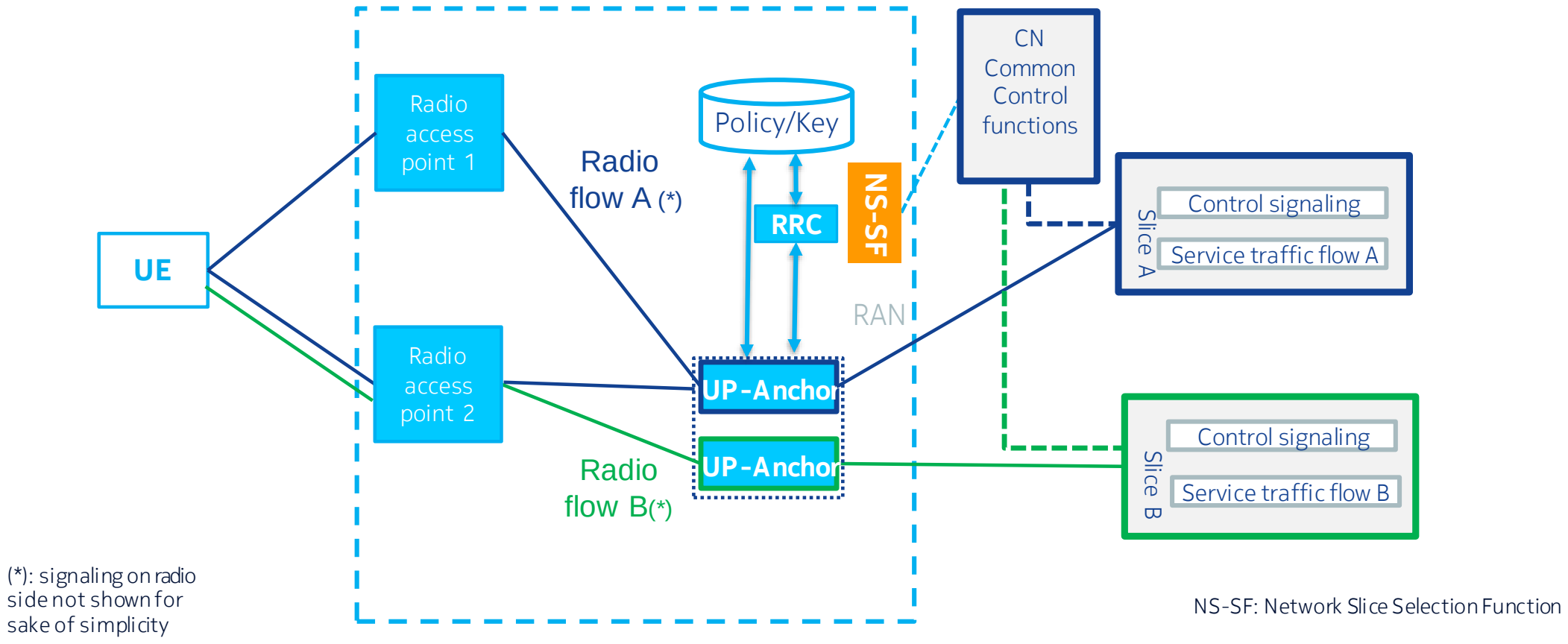

**Figure 3: Multi-connectivity anchor – The interface between Network slicing and RAN**

Radio resource management and control in the base station, and correspondingly in the UE if several slices are configured, is responsible for configuring the RAN protocol stack and QoS according to the slice requirements. For example, for a slice with high throughput requirements, radio bearers are configured to support multi-connectivity (MC), e.g., similar to the split bearer approach as in LTE dual connectivity or in the equivalent in 5G. For slices with low-latency and high robustness requirements, lower frame error rates as well as multi-point diversity techniques may be utilized.

In the example illustrated in Figure 3, the radio flow in network slice A, which could correspond to a radio bearer in LTE, is configured with two radio connections, while network slice B is configured with only one connection according to the provided policy configuration. In summary, network slicing can be realized by appropriate mapping control and configuration of radio network functions without changing fundamental paradigms of the RAN.

### *B. Multiplexing network slices in RAN*

The RAN is a typical example of a shared network function controlled by a single authority, where spectrum is shared amongst mobile virtual network and service operators. Figure 4 illustrates an example of a common spectrum shared by three network slices, each with own RAN and CN part. The layer 2 Control-plane is split into cell related functions which are common to all slices, and session or user specific radio resource control (RRC). Depending on the underlying service, RRC can configure and tailor the User-plane protocol stack. For example, for a slice supporting low delay services IP and related Header Compression (HC) may not be used, and RLC can be configured in transparent mode. In contrast, for services requiring QoE and excellent QoS, IP as well as acknowledged RLC must be initiated. In addition, there would be the possibility to chain proprietary and operator specific functions within a network slice. In this regard, the intra-slice application scheduler (which prioritizes sessions within the related slice) is chained in RAN slice 1, while the inter-slice radio scheduler (which schedules different slices) resides in the common RAN part (c.f. Figure 4) and makes use of multi-service scheduling capabilities. Multi-service scheduling is part of a flexible RAN and provides the capabilities to differentiate traffic classes and to assign resources according to QoS requirements. Hence, service flows from different slices can be individually treated, e.g., flexible numerologies can be used to fulfil QoS constraints and even semi-persistently reserved resources for deterministic traffic requirements.

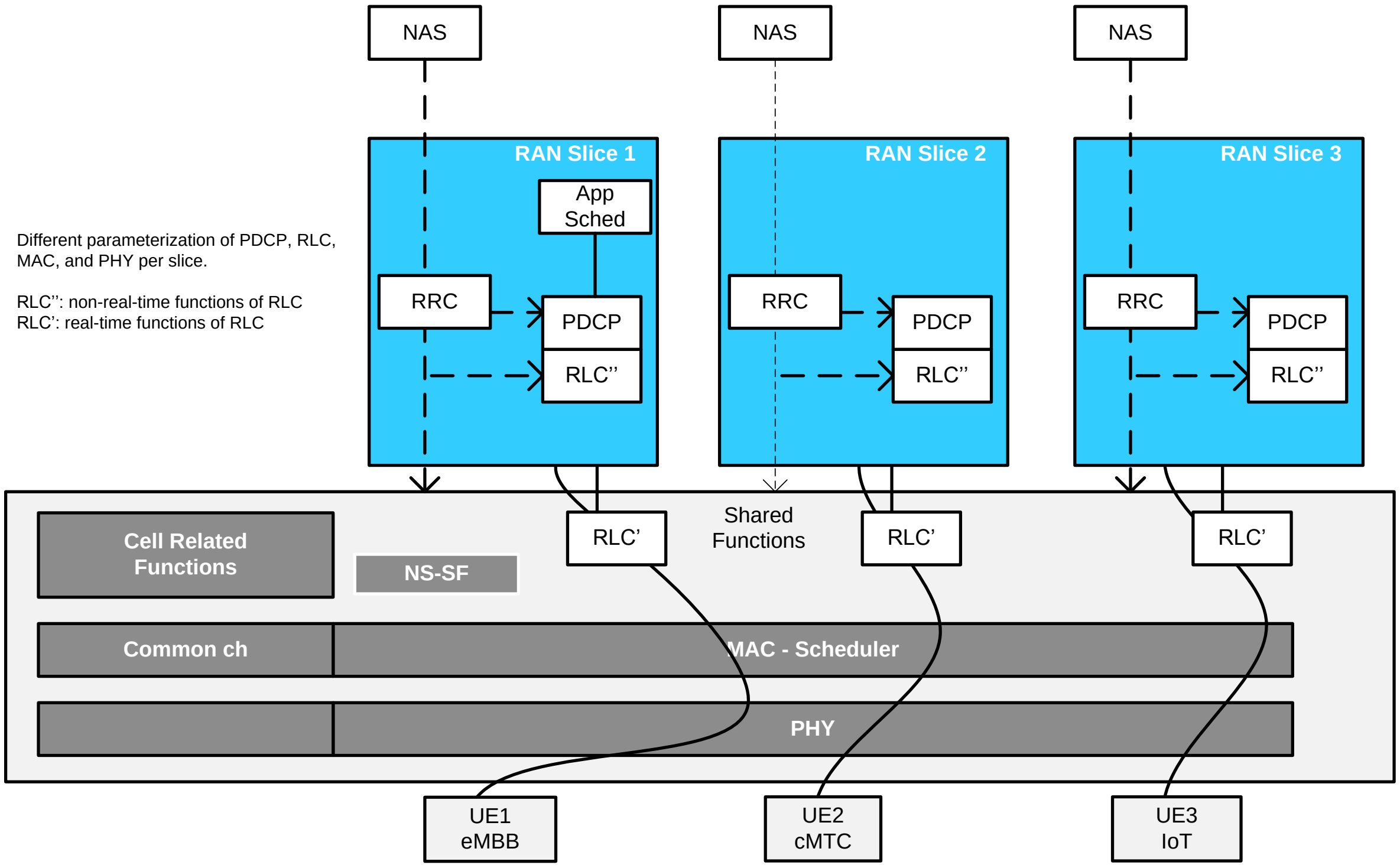

**Figure 4: Example of common spectrum shared by multiple slices**

It is worth mentioning that, although the higher RAN layers can be configured to operate in a slice-specific mode relatively easily, this is not the case for the lower-layer radio interface. In contrast to current 3GPP LTE, where radio slices are represented by new variants of 3GPP such as Narrowband-IOT (NB-IOT), 5G requires the inherent coexistence of diverse services. Hence, in contrast to 4G LTE where adding a new radio slice requires modification to the legacy LTE radio, the new radio proposed for 5G is designed to be forward compatible [11], among others by utilizing new radio framing and protocols. This means that future addition of new services and thus radio slices will not require changes in the 5G radio framework.

In a similar context, it is worth pointing out that the new radio framing involves the so-called "tiling" concept [12]. That is, time and frequency resources of the new 5G radio are tiled so that it can be allocated for the needs of certain slices with given requirements. An illustration of the tiling concept is provided in Figure 5.

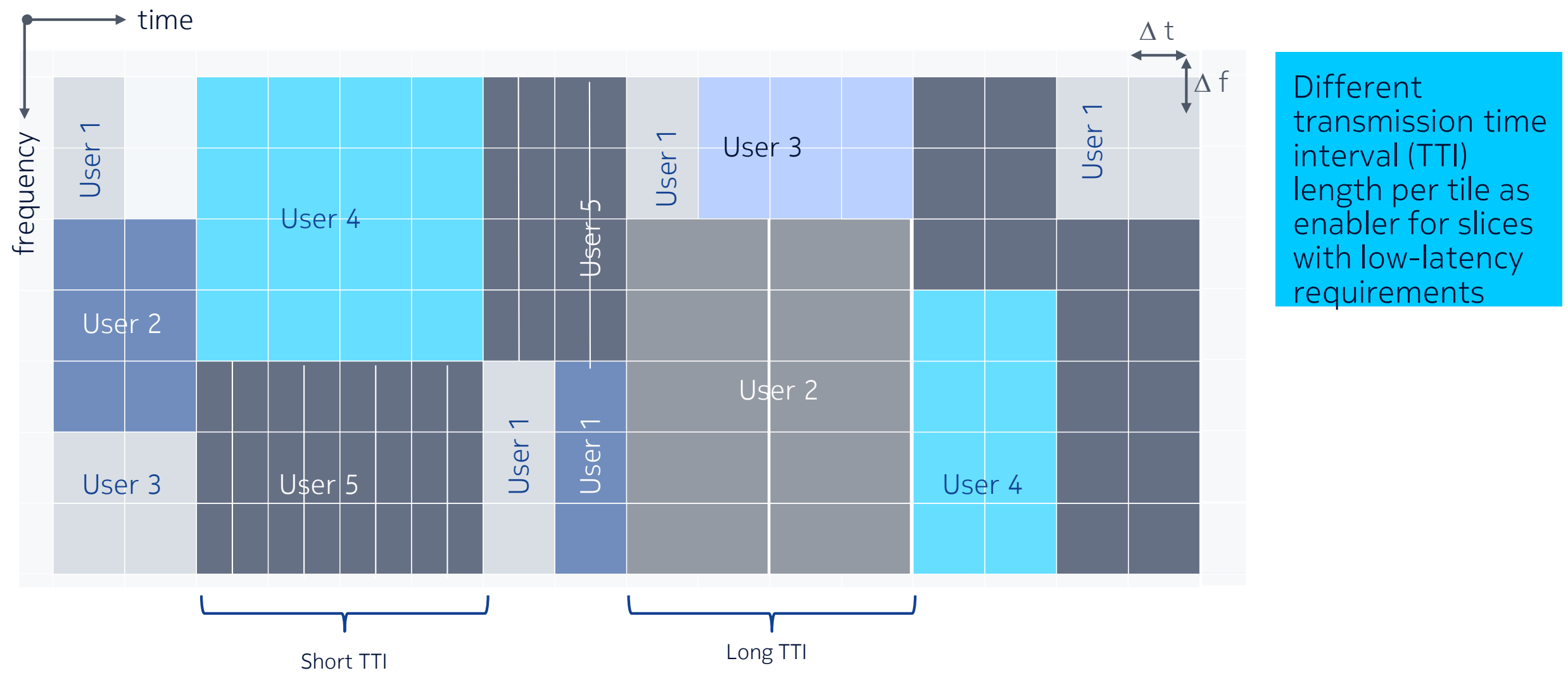

**Figure 5: The radio tiling concept proposed for 5G radio with network slicing**

### *C. Exemplary architecture with shared RAN slices*

An exemplary architecture with shared RAN slices is presented in Figure 6a), which shows how the different aspects may integrate, based on three options [13]:

1) The first option (Option 1) shows two network slices where each slice carries two different services. Each slice may be operated by a different mobile network operator (MNO). Furthermore, for each slice an individual RAN protocol stack is implemented down to the upper part of the physical layer. Only the lower part of the physical layer is shared across slices. The multiplexed access to the transponder part of the physical layer (PHY-TP) is coordinated by the SDM-X which makes use of flexible and efficient radio resource management for supporting different numerologies within the same spectrum. One could think of Option 1 as implementing all user-specific functions such as forward error correction encoding, layer mapping and precoding in an individual fashion, while TP-specific functionality such as transmission of synchronization and cell-specific reference signals are shared.

2) Option 2 depicts again two network slices from two operators. Compared to the previous example, each slice uses an individual implementation of service-specific functionality such as PDCP, RLC, and slice-specific RRC. In addition, the tenant may implement a customized QoS scheduling to perform pre-scheduling. The access to the MAC layer is then controlled by the SDM-X where resource fairness across tenants and QoS guarantees corresponding to individual SLAs must be met. Furthermore, resource isolation must be provided to alleviate side-effects.

3) Option 3 illustrates the case of two operators using the same RAN as shared resource, i.e., the SDM-X is the interface between CN and RAN. In this example, no customization of radio resource management beyond SDM-X parameters and configuration would be possible.

### *D. Flexible RAN Technologies as Enablers for Shared RAN*

In addition to the flexible architecture considerations mentioned above, further flexible RAN technologies enable a shared RAN for network slicing and accommodating highly diverse services. In the following, some of them are briefly explained and how they facilitate network slicing.

**Multi-connectivity (MC):** The term RAN MC refers to the versatile scenario where a UE connects to the network via multiple cells. For the sake of the current explanation, it suffices to consider that a multi-connectivity approach takes place whenever the connection of the UE to the RAN involves multiple PHY interfaces. Those multiple PHY interfaces are leveraged to deliver enhanced performance capabilities, which are translated into aggregated throughput or increased reliability. A major challenge is to enforce different QoS requirements, differentiation, and prioritization within a RAN exploiting MC and Multi-RAT through a single scheduler.

Next, we consider two MC options, namely the *common PDCP* and *common MAC approach*, which are shown in Figure 6b), as part of the exemplary architecture, Option 2, discussed in Section III-C. The *common PDCP approach* dictates that the PDCP layer of the protocol stack is shared between the individual connections of the RAN multi-connectivity (henceforth called "radio leg"), and all layers below PDCP are separate logical entities. Such approach resembles that of dual connectivity in 3GPP LTE, and offers the advantage of flexibility in terms of the physical location of the protocol stack layers. The main advantage of the common PDCP approach is the flexibility it offers in terms of the physical location of the protocol stack layers. In particular, since the interface between PDCP and RLC is not a time-critical interface, the common PDCP layer is not necessarily co-located with RLC, hence mobility-related signaling can be hidden from the CN. In the *common MAC approach*, the multi-connectivity anchor point is the MAC layer, similarly as carrier aggregation in 3GPP LTE. Owing to the time-critical interface between MAC and PHY, the common MAC approach requires that either the multi-connectivity legs originate from the same site, or they are interconnected via a high-capacity transport link. Nevertheless, the common MAC approach offers the advantage of fast information exchange between the different multi-connectivity legs. This facilitates coordinated scheduling, interference mitigation, and other schemes related to MAC scheduling [14].

**Multi-RAT and millimeter wave (mmW) technology:** It is envisioned that mmW technology will play a key role in the fulfillment of 5G network requirements. MC will be essentially required to support mmW deployments, which are anticipated to cover both mobile broadband and machine-type applications. The design characteristics of such deployments will depend upon factors which span a wide area of architecture requirements, such as transport capabilities, low-band integration, propagation impairments and (edge or core) cloud implementations. Consequently, a flexible architecture incorporating mmW support is required to meet different slice requirements.

**User-centric signaling:** A user centric signaling and mobility management for services including short, sporadic and delay tolerant data packets is proposed based on a User Centric Connection Area (UCA) [15]. The UCA consists of a set of radio nodes selected by the flexible 5G-RAN. One radio node acts as an anchor node within the UCA, which shares the user-context with all other nodes within the UCA. The CN connections (bearers) are terminated at the anchor node. With the help of a shared context, mobility is managed by the RAN instead of the CN as long as the UE moves within the UCA. This implies that mobility is hidden towards the core network, which reduces mobility and connection related signaling. Based on the context sharing, the UE is able to send UL packets and receive DL packets by any node within a UCA. The user specific aspect provides a flexibility and re-configurability in the realization of an UCA, i.e., each UCA can be configured according to specific requirements taking into account QoS parameters.

**Mobile edge computing and edge cloud processing:** Advanced 5G services are envisioned to be offered at the network edge so as to reside much closer to the user in order to enhance delay and perceived performance, e.g., adopting the ETSI MEC paradigm[1]. Therefore, a flexible service chaining should also be improved to establish dynamic services considering edge network locations and might be combined with VNFs to ensure a joint optimization of services and networking operations. Edge server locations can also be exploited for storage, computation, and dynamic service creation within a given network slice by verticals and over-the-top providers, introducing another multi-tenancy dimension.

[1] http://www.etsi.org/technologies-clusters/technologies/mobile-edge-computing

## IV. Conclusions and Further Challenges

An overview of the basic implementation features of network slicing was presented, along with its potential to provide revenue to the network operator. The analysis included the basic principles behind the mapping of dedicated and shared slices, as well as implementation-specific aspects when the concept of network slicing is employed over RAN and CN. Special focus was put on the connection of network slicing with RAN concepts. Based on the above analysis, a strong potential of network slicing was revealed for addressing the diverse requirements of future 5G systems. Nevertheless, network slicing remains still at an early stage in terms of its development, hence one should anticipate a long way before it becomes a mature technology and thus be adopted by network standards.

## V. Acknowledgement

This work has been performed in the framework of the H2020-ICT-2014-2 project 5G NORMA. The authors would like to acknowledge the contributions of their colleagues. This information reflects the consortium's view, but the consortium is not liable for any use that may be made of any of the information contained therein.

## VII. Biographies

**Peter Rost** [SM] (peter.m.rost@nokia-bell-labs.com) received his Ph.D. degree from Technische Universität Dresden in 2009 his M.Sc. degree from University of Stuttgart in 2005. Since May 2015, Peter is member of the Radio Systems research group at Nokia Germany, contributing to the European projects 5G-NORMA and METIS-II, and business unit projects on 5G Architecture. Peter serves as member of IEEE ComSoc GITC, VDE ITG Expert Committee Information and System Theory, and as Executive Editor of IEEE Transactions of Wireless Communications.

**Christian Mannweiler** (christian.mannweiler@nokia-bell-labs.com) received his M.Sc. (Dipl.-Ing.) and Ph.D. (Dr.-Ing.) degrees from University of Kaiserslautern (Germany) in 2008 and 2014, respectively. Since 2015, he is a member of the Cognitive Network Management research group at Nokia Bell Labs Germany, where he has been working in the area of network management automation and SON for 5G systems. Christian contributed to several nationally and EU-funded projects covering architecture design of mobile communication systems, among them 5G NORMA.

**Diomidis S. Michalopoulos** [SM] (diomidis.michalopoulos@nokia-bell-labs.com) received his PhD from the Aristotle University, Greece, in 2009. From 2009 to 2015 he was UBC university, Canada, as Killam and Banting postdoctoral fellow, and FAU university, Germany, as researcher and teaching instructor. In 2015 he joined Nokia Bell Labs, Germany, contributing towards radio and architecture aspects of 5G. Diomidis received several awards for academic achievement including the Marconi Young Scholar Award, and served as Editor of the IEEE Communications Letters.

**Cinzia Sartori** (cinzia.sartori@nokia-bell-labs.com) is a principal expert in the field of end-to-end Mobile Network Architecture with focus on 5G at NOKIA in Munich. Until mid 2013 she led the ‘Self-Organizing Network (SON) Research and Standardization’ project in Nokia Siemens Networks. She is co-author of the ‘LTE Self-Organizing Network’ book (January 2012). Earlier she worked in the Network Telecom, O&M, RRM and SS7 in Nokia Siemens Networks, Siemens and GTE. She graduated as engineer in Pavia -Italy.

**Vincenzo Sciancalepore** [S’11–M’15] (vincenzo.sciancalepore@neclab.eu) received the M.Sc. degree in telecommunications engineering and telematics engineering in 2011 and 2012, respectively, and the double Ph.D. degree in 2015. From 2011 to 2015, he was a Research Assistant with IMDEA Networks, where he was involved in intercell coordinated scheduling for LTE-Advanced networks and device-to-device communication. He is currently a Research Scientist with NEC Europe Ltd., Heidelberg, where he is involved in network virtualization and network slicing challenges.

**Nishanth Sastry** (nishanth.sastry@kcl.ac.uk) is a Senior Lecturer at King's College London. He holds a PhD from the University of Cambridge, UK, a Master's degree from The University of Texas at Austin, and a Bachelor's degree from Bangalore University, India, all in Computer Science. He has over six years of experience in the Industry (Cisco Systems, India and IBM Software Group, USA) and Industrial Research Labs (IBM TJ Watson Research Center).

**Oliver Holland** [M] (oliver.holland@kcl.ac.uk) is a senior researcher at King's College London, working on spectrum sharing and 5G technologies among others. He has an established record in various standards activities, including leadership of various standards, and has undertaken project, academic, event and other leaderships, within the IEEE and elsewhere. He has achieved various accolades for his work, most recently creating and leading a team/submission/idea that won the 0.5m EUR European Union Collaborative Spectrum Sharing Prize.

**Shreya Tayade** (tayade@eit.uni-kl.de) received her B.E. from Government College of Engineering Aurangabad, India in 2010 and M.Sc. degree in Communication Engineering from RWTH Aachen University, Germany in 2015. Since 2015, she is a researcher in Technical University of Kaiserslautern Germany and is working on 5G based EU and industrial projects.

**Bin Han** (binhan@eit.uni-kl.de) received the B.E. (2009) and M. Sc. (2012) degrees from the Shanghai Jiao Tong University and the Technische Universität Darmstadt, respectively. In 2016 he was granted the Ph.D. (Dr.-Ing.) degree in Electrical Engineering and Information Technology from the Karlsruhe Institute of Technology. He is currently a postdoctoral researcher at the University of Kaiserslautern, researching in the broad area of communication systems and signal processing, with a special focus on 5G.

**Dario Bega (**dario.bega@imdea.org) received his B.Sc. and M.Sc. degree in Telecommunication Engineering in 2010 and 2013 respectively, from University of Pisa, Italy. Since 2015, he is a Research Assistant with Imdea Networks in Madrid and a PhD student in Telematic Engineering with Universidad Carlos III de Madrid. His research work focuses in Wireless Network and Multi-tenancy approach for 5G.

**Danish Aziz** (danish.aziz@nokia-bell-labs.com) is a research engineer at Nokia Bell Labs Germany. He received his M.Sc. and Ph.D. (Dr.-Ing.) degrees in Communications Engineering from University of Stuttgart, Germany. He started his career in 2006 with the development of 4G-LTE systems. He is currently involved in the development flexible radio access network architecture for 5G. Danish has worked in several EU-funded projects e.g. METIS-I. He is actively contributing in ongoing 5G-NORMA project.

**Hajo Bakker (**hajo.bakker@nokia-bell-labs.com**)** is team leader within End-To-End Mobile Networks Research at Nokia Bell Labs, Stuttgart, Germany. Hajo received his Dipl.-Ing. degree in electrical engineering telecommunication from the Technical University of Brunswick in Germany and joined Alcatel research in 1986. He participated in the 3GPP RAN 2 working group for the standardization of LTE. Hajo is the author/co-author of more than 20 scientific publications, and holds more than 80 granted patents and patent applications.